\documentclass[manuscript]{aastex62}
\usepackage{natbib}
\received{}
\revised{}
\accepted{}
\submitjournal{ApJ}

\shorttitle{[\ion{Fe}{2}] and H$_2$ line mapping of IC~443}
\shortauthors{Kokusho et al.}
\begin{document}

\title{Near-infrared [\ion{Fe}{2}] and H$_2$ line mapping of the supernova remnant IC~443 with the IRSF/SIRIUS}

\correspondingauthor{Takuma Kokusho}
\email{kokusho@u.phys.nagoya-u.ac.jp}

\affil{Graduate School of Science, Nagoya University, Chikusa-ku, Nagoya 464-8602, Japan}

\author{Takuma Kokusho}
\affil{Graduate School of Science, Nagoya University, Chikusa-ku, Nagoya 464-8602, Japan}

\author{Hiroki Torii}
\affil{Graduate School of Science, Nagoya University, Chikusa-ku, Nagoya 464-8602, Japan}

\author{Takahiro Nagayama}
\affil{Department of Astrophysics, Kagoshima University, 1-21-35 Korimoto, Kagoshima, 890-0065, Japan}

\author{Hidehiro Kaneda}
\affil{Graduate School of Science, Nagoya University, Chikusa-ku, Nagoya 464-8602, Japan}

\author{Hidetoshi Sano}
\affil{Graduate School of Science, Nagoya University, Chikusa-ku, Nagoya 464-8602, Japan}

\author{Daisuke Ishihara}
\affil{Institute of Space and Astronautical Science, Japan Aerospace Exploration Agency, 3-1-1 Yoshinodai, Chuo-ku, Sagamihara, Kanagawa 252-5210, Japan}

%\author{Ho-Gyu Lee}
%\affil{Korea Astronomy and Space Science Institute, Daejeon 305-348, Republic of Korea}

\author{Takashi Onaka}
\affil{Department of Physics, Faculty of Science and Engineering, Meisei University, Hodokubo, Hino, Tokyo 191-8506, Japan}
\affil{Department of Astronomy, Graduate School of Science, The University of Tokyo, Bunkyo-ku, Tokyo 113-0033, Japan}



\begin{abstract}
We investigate properties of the interstellar medium (ISM) interacting with shocks around the Galactic supernova remnant IC~443, using the results of near-infrared [\ion{Fe}{2}] and H$_2$ line mapping with the IRSF/SIRIUS. In the present study, we newly performed H$_2$ 1--0 S(1) and 2--1 S(1) line mapping with the narrow-band filters tuned for these lines, covering the entire remnant ($30{\arcmin}\ {\times}\ 35{\arcmin}$). Combined with an [\ion{Fe}{2}] line map, our result shows that the H$_2$ line emission is significantly detected in the southern region, while the [\ion{Fe}{2}] line emission is detected all over the remnant, suggesting that slow and fast shocks propagate through the southern region and the entire remnant, respectively. In particular, the H$_2$ line emission is relatively strong compared to the [\ion{Fe}{2}] line emission in the southwestern region, where TeV $\gamma$-ray emission is detected. As the strong H$_2$ line emission indicates the dominance of the dense ISM, this result supports the scenario that the $\gamma$-ray emission is likely to be produced through a heavy interaction between cosmic-ray protons and the dense ISM in the southwestern shell. We also find that the H$_2$ and [\ion{Fe}{2}] line emissions show an anti-correlated spatial distribution in the same region, suggesting the presence of the clumpy ISM. Such a clumpy morphology of the ISM around IC~443 may assist cosmic-ray protons to efficiently interact with large amounts of the ISM protons.
\end{abstract}

\keywords{infrared: ISM --- ISM: individual objects (IC~443) --- 
ISM: supernova remnants --- (ISM:) cosmic rays}


\section{Introduction} \label{sec:intro}

Massive stars explode at the end of their lives as a supernova, producing shock waves and releasing heavy elements into the interstellar space. Cosmic rays are assumed to be accelerated by supernovae, where protons and electrons gain high energy up to ${\sim}10^{15}$ eV in shock fronts by diffusive shock acceleration \citep[e.g.][]{bel78,aha07}. Some evidence for cosmic ray acceleration by supernovae has been confirmed through observations of supernova remnants (SNRs) in X-ray and $\gamma$-ray emissions. For instance, hard synchrotron X-ray emission due to cosmic-ray electrons is detected from SN~1006 and Cas~A \citep{koy95,all97}. The $Fermi$ satellite confirms that the $\gamma$-ray emission from IC~443 and W~44 is produced in the decay of neutral pions originating via collisions between cosmic-ray protons and protons in the interstellar medium \citep[ISM;][]{ack13}. In order to produce strong $\gamma$-ray emission of cosmic-ray proton origins, a large amount of the ISM protons should be present near SNRs as targets of cosmic-ray protons. From this point of view, for the TeV $\gamma$-ray SNR RX~J1717.7--3946, \citet{fuk12} compared spatial distributions of atomic and molecular protons obtained from radio observations with that of the TeV $\gamma$ rays, demonstrating spatial coincidence of the ISM protons with the $\gamma$-ray emission. Similarly, \citet{yos13} show spatial coincidence of the ISM protons with $\gamma$-ray emission in W~44.

IC~443 is a Galactic SNR located towards the Galactic anti-center, with the age estimated to be $3,000$--$30,000$ yr \citep{pet88,olb01}. IC~443 is known to heavily interact with the surrounding ISM, suggesting that cosmic-ray protons may be interacting with a large amount of the ISM protons. Hence IC~443 is expected to be bright in the $\gamma$-ray emission, and indeed several $\gamma$-ray observations with both ground-based and space telescopes reveal that IC~443 is bright in GeV and TeV $\gamma$ rays of cosmic-ray proton origins \citep{alb07,acc09,tav10,ack13}. And yet acceleration mechanisms of cosmic rays are not fully understood. In order to address this issue, more detailed observations and analyses of $\gamma$-ray spectra of SNRs are essential. However $\gamma$-ray spectra from SNRs are hard to interpret because of the effects of the ISM around SNRs. For instance, \citet{ino10} suggest that the $\gamma$-ray spectrum of cosmic-ray origins can be a broken power-law spectrum due to cosmic-ray acceleration in multiple reflected shocks when shocks interact with the clumpy ISM. In addition, such clumpy and thus dense clouds interacting with shocks are thought to promote cosmic-ray acceleration through magnetic field amplification and/or magnetic reconnection in the turbulent medium \citep[e.g.][]{ino12,san15,mar18}. Hence, to understand acceleration mechanisms of cosmic rays, we need to investigate morphologies of the ISM interacting with shocks in detail. Molecular clouds surrounding IC~443 are first discovered by \citet{cor77}, and many studies have been investigated detailed distributions and properties of the molecular clouds associated with IC~443 \citep[e.g.][and references therein]{su14}. Yoshiike et al. (2020, in prep.) recently demonstrate that the ISM shows an excellent spatial coincidence between the distribution of $\gamma$ rays and that of molecular gas in IC~443, whereas the spatial resolution of the molecular gas observation may not be high enough to assess detailed morphologies of the ISM.

In addition to radio observations of atomic and molecular gas emissions, infrared (IR) observations also enable us to study properties of the ISM surrounding SNRs, tracing atomic gas, molecular gas, and dust interacting with shocks. In particular, near-IR [\ion{Fe}{2}] and H$_2$ line emissions are important probes to investigate the shocked ISM in SNRs. In general, most Fe is locked up in dust grains \citep[e.g.][]{dra95}. In shock regions, however, dust is destroyed and accordingly gas-phase Fe is released into the interstellar space, producing [\ion{Fe}{2}] line emission in the near- and mid-IR wavelengths. Because fast shocks (${\ga}150$ km s$^{-1}$) are needed to destroy dust grains \citep{jon94}, [\ion{Fe}{2}] line emission traces the low-density shocked ISM where shocks are not strongly decelerated. In contrast, H$_2$ line emission is bright in slow-shock regions where dust is not heavily destroyed, because shocks can heat but not dissociate H$_2$ at a velocity ${\la}50$ km s$^{-1}$ \citep{dra83}. Hence H$_2$ line emission traces the dense shocked ISM where shocks are strongly decelerated. Combined with a relatively high spatial resolution of near-IR observations, these unique properties of the near-IR [\ion{Fe}{2}] and H$_2$ line emissions allow us to assess physical properties including morphologies of the ISM around SNRs in detail. Some studies have found that IC~443 is bright in the near-IR [\ion{Fe}{2}] and H$_2$ line emissions \citep[e.g.][]{tre79,gra87}, suggesting that both fast and slow shocks are interacting with the ISM in the SNR. \citet{all93} discuss the properties of the shocked ISM in a herbig-halo object utilizing differences in the spatial distribution between [\ion{Fe}{2}] and H$_2$ line emissions, while no one has fully conducted a similar study on IC~443 so far.

In the present study, we report results of mapping observations of IC~443 in the near-IR [\ion{Fe}{2}] and H$_2$ line emissions. $\gamma$-ray emission due to cosmic-ray protons is detected around the southwestern region of the SNR \citep{alb07,acc09,abd10}, where molecular gas is also detected \citep[e.g.][]{lee12}. To reveal morphologies of the ISM interacting with shocks, we study spatial distributions of the [\ion{Fe}{2}] and H$_2$ line emissions. As the [\ion{Fe}{2}] line intensity maps are presented in our previous studies \citep{kok13,kok15}, we newly present the H$_2$ line intensity maps in the present study. The details of our observation and data reduction are presented in Sect.~\ref{sec:obs}, and our results in Sect.~\ref{sec:res}. In Sect.~\ref{sec:dis}, we discuss properties of the ISM interacting with shocks in IC~443, and we present our conclusion in Sect.~\ref{sec:con}.  IC~443 is located in the Galactic anti-center at a distance of $1.5$ kpc \citep{wel03}, which is used throughout the paper.

\section{Observations and data reduction} \label{sec:obs}

We performed [\ion{Fe}{2}] and H$_2$ line mapping of IC~443 with the near-IR camera SIRIUS \citep[Simultaneous Infrared Imager for Unbiased Survey;][]{nag99,nag03} installed on the IRSF (InfraRed Survey Facility) 1.4-m telescope, located at the South African Astronomical Observatory, using the narrow-band filters tuned for [\ion{Fe}{2}] 1.257~{\micron}, [\ion{Fe}{2}] 1.644~{\micron}, H$_2$ 1--0 S(1), and H$_2$ 2--1 S(1). The SIRIUS camera has a field of view and a pixel scale of $7{\farcm}7\ {\times}\ 7{\farcm}7$ and $0{\farcs}45$, respectively. The seeing size during the observation was typically $1{\arcsec}$. We observed $44$ overlapping fields to cover the entire remnant ($30{\arcmin}\ {\times}\ 35{\arcmin}$). The details of the observations including the bandpasses of the filters of the [\ion{Fe}{2}] line emission are described in \citet{kok13}, while those of the H$_2$ line emission are summarized in Table.~\ref{table:obs}.

In order to create the H$_2$ line intensity maps, we first applied the same standard procedure as in \citet{kok13}. Then we newly applied spatial high-pass filtering to the H$_2$ line intensity maps because the intensity of the map is affected by a thermal emission of the instrument as well as by OH line emissions. In the filtering procedure, we adopted a cut-off spatial frequency of $1{\arcmin}$ after masking the regions where the H$_2$ 1--0 S(1) line emission was detected with signal-to-noise ratios ($S/N$) higher than $1.5$. We performed the flux calibration of the H$_2$ line intensity maps using the $K$-band magnitudes of stars from the Two Micron All Sky Survey (2MASS) Point Source Catalog \citep{skr06} and assuming that the magnitudes of stars are the same between the narrow- and broad-band images. The number of the stars used for the photometric calibration is typically $10$ per image, and the uncertainty of the calibration coefficient is typically $3{\%}$. The calibration coefficient can also vary with ${\leq}5{\%}$ if we assume a blackbody spectrum in the photometric calibration. We masked the point sources with SExtractor \citep{ber96} not to overestimate the [\ion{Fe}{2}] and H$_2$ line intensities due to the stellar emission. Extinction correction was not performed to the [\ion{Fe}{2}] and H$_2$ line intensities.

\section{Results} \label{sec:res}

\begin{figure}
\plotone{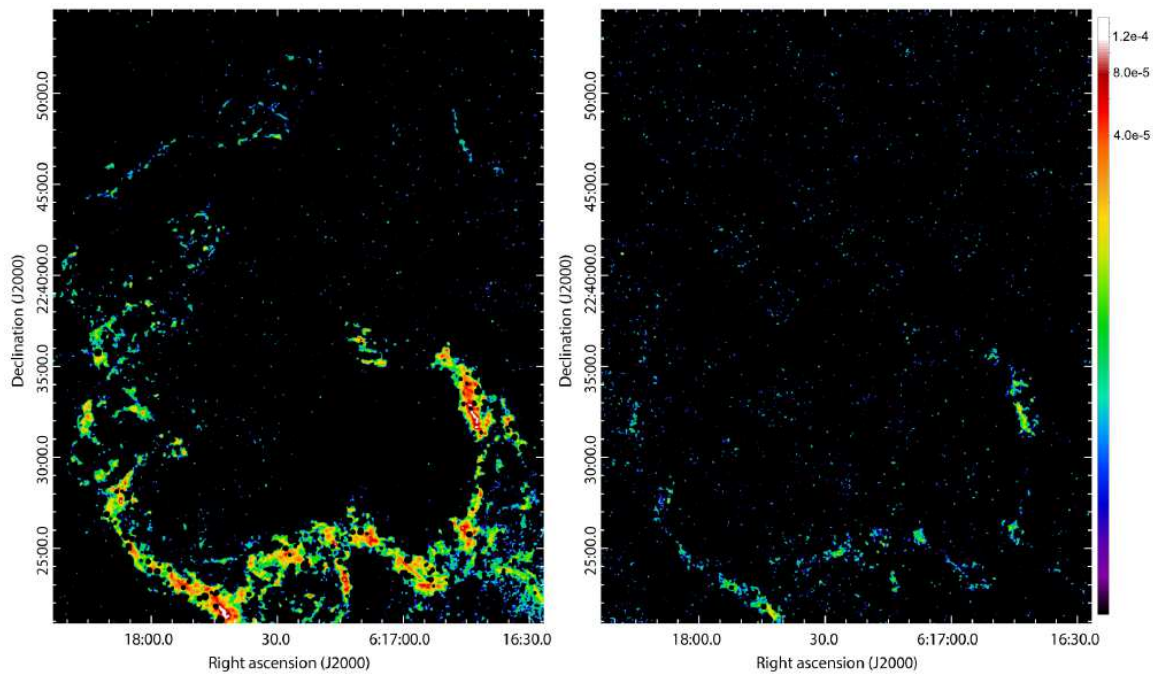}
\caption{H$_2$ 1--0 S(1) (left) and 2--1 S(1) (right) line intensity maps, where point sources are removed. The maps are smoothed with a Gaussian kernel of $2{\farcs}3$ in sigma. The color levels are given in units of ${\rm ergs}\ {\rm s}^{-1}\ {\rm cm}^{-2}\ {\rm sr}^{-1}$.\label{fig:h2}}
\end{figure}

Figure~\ref{fig:h2} shows the H$_2$ 1--0 S(1) and 2--1 S(1) line intensity maps of IC~443. For the H$_2$ brightest position at $({\alpha}_{\rm J1950}$, ${\delta}_{\rm J1950})$=$(6^{\rm h}14^{\rm m}41{\fs}7$, +$22{\arcdeg}22{\arcmin}40{\arcsec})$, \citet{bur88} measured the H$_2$ 1--0 S(1) and 2--1 S(1) line fluxes in a $19{\farcs}6$ aperture to be $(3.1{\pm}0.6){\times}10^{-12}$ and $(2.4{\pm}0.7){\times}10^{-13}$~erg~s$^{-1}$~cm$^{-2}$, respectively. We measured the H$_2$ 1--0 S(1) and 2--1 S(1) line fluxes at the same position with our maps to be $(2.8{\pm}0.1){\times}10^{-12}$ and $(2.9{\pm}0.1){\times}10^{-13}$~erg~s$^{-1}$~cm$^{-2}$, respectively, confirming that our line fluxes are consistent with those of \citet{bur88}.  The largest map of the near-IR H$_2$ line emission of IC~443 so far was presented by \citet{bur90} with a field of view of $20{\arcmin}\ {\times}\ 24{\arcmin}$. We improve the H$_2$ line intensity map to cover the entire remnant ($30{\arcmin}\ {\times}\ 35{\arcmin}$), showing that the H$_2$ line emission is bright in the southern region as revealed by previous studies \citep[e.g.][]{bur88,rho01}. In the present study, we further reveal that faint H$_2$ line emission is present in parts of the northeastern shell. \citet{rho01} predicted that the $K$-band emission in the same region may be dominated by Br$\gamma$ from the 2MASS observations, based on the detection of Br$\gamma$ by \citet{gra87} in the northeastern shell.

Figure~\ref{fig:h2-fe} shows the pseudo-color [\ion{Fe}{2}] $1.644~{\micron}$ and H$_2$ 1--0 S(1) line intensity maps of IC~443, with contours of the $^{12}$CO ($J=1$--$0$) line emission (Yoshiike et al. 2020, in prep.) and the TeV $\gamma$-ray emission \citep{acc09} obtained from observations with the NANTEN2 and VERITAS telescopes, respectively. The figure demonstrates that the [\ion{Fe}{2}] line emission is detected all over the region, indicating that fast shocks propagate in the entire remnant. On the other hand, the figure shows that the H$_2$ line emission is significantly bright in the southern region, suggesting that slow shocks mainly propagate in the southern shell. This also suggests that the southern shell is dominated by the dense ISM, which is likely to decelerate shocks. Such a picture has been indicated by previous studies on the shocked ISM in the southern region of the SNR \citep[e.g.][]{bur88,ric95,rho01}. Indeed, as can be seen from the figure, molecular clouds dominate the southern shell, showing an excellent spatial agreement in the distributions between CO and H$_2$ emissions. Higher spatial-resolution observations have also revealed resemblance in the spatial distribution between CO and H$_2$ emissions around $({\alpha}_{\rm J1950}$, ${\delta}_{\rm J1950})$=$(6^{\rm h}14^{\rm m}41{\fs}7$, +$22{\arcdeg}22{\arcmin}40{\arcsec})$  \citep{wan92}. On the other hand, the [\ion{Fe}{2}] line emission tends to be spatially separated from the CO emission. In particular, the H$_2$ line emission is relatively bright as compared to the [\ion{Fe}{2}] line emission in the southwestern region, where the TeV $\gamma$ rays are detected. It is also notable from the figure that the [\ion{Fe}{2}] shell is distributed outside the H$_2$ shell in the southwestern region, while the [\ion{Fe}{2}] shell is located inside the H$_2$ shell in the southeastern region. This spatial difference between the [\ion{Fe}{2}] and H$_2$ shells will be discussed later.

\begin{figure}
\centering
\includegraphics[scale=0.5]{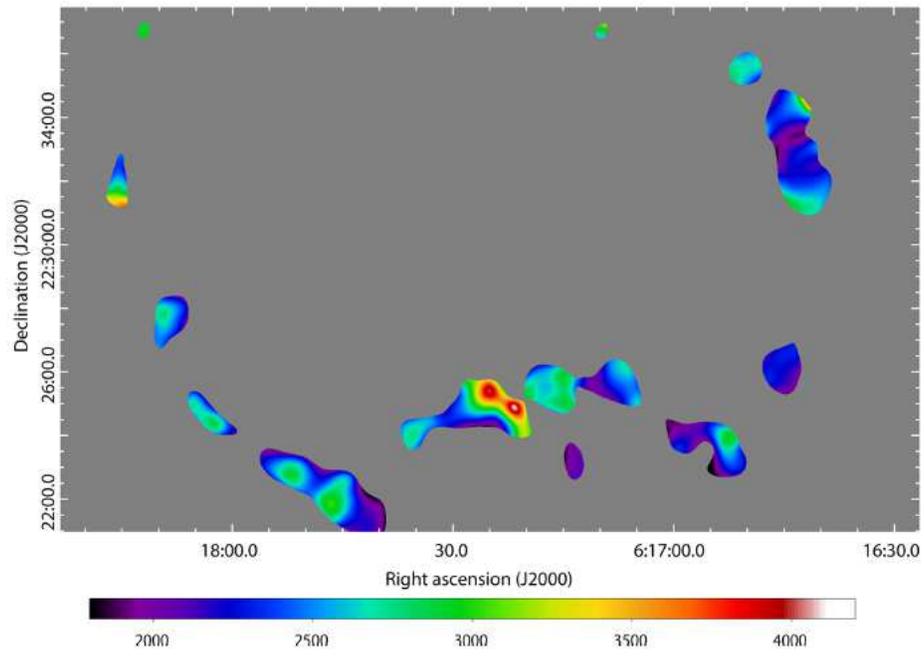}
\caption{Distribution of the H$_2$ excitation temperature derived from the H$_2$ 2--1 S(1)/1--0 S(1) line ratio. The color levels are given in units of K.
\label{fig:h2temp}}
\end{figure}

Figure~\ref{fig:h2temp} shows the distribution of the H$_2$ excitation temperature calculated from the H$_2$ 1--0 S(1)/2--1 S(1) line ratio. To create this, we first smoothed the H$_2$ line intensity maps with a Gaussian kernel of $18{\arcsec}$ in sigma. Then we masked the regions where the H$_2$ 1--0 S(1) and 2--1 S(1) line emissions are not detected with $S/N>5$. Here we adopted the transition probabilities presented by \citet{tur77}. Except for the southern edge region (see below), the resultant H$_2$ excitation temperature is around $2000$~K, which is known to be typical for shocked molecular gas in SNRs \citep[e.g.][]{ric95}. Since a large spatial resolution element can contain many separate shocks, we evaluate the uncertainty of the temperature in Figure~\ref{fig:h2temp}, using the maps smoothed with a Gaussian kernel of $2{\farcs}3$ in sigma (i.e. Figure~\ref{fig:h2}). In these maps, the H$_2$ 1--0 S(1) and 2--1 S(1) line emissions are detected with $S/N>5$ around $({\alpha}_{\rm J2000}$, ${\delta}_{\rm J2000})$=$(06^{\rm h}17^{\rm m}43^{\rm s}, +22{\degr}21{\arcmin}28{\arcsec})$ and $(06^{\rm h}16^{\rm m}43^{\rm s}, +22{\degr}32{\arcmin}02{\arcsec})$, where the standard deviation of the temperature within the spatial resolution of Figure~\ref{fig:h2temp} is estimated to be ${\sim}200$~K. As the temperature of shocked gas is a measure of shock velocities for gas densities of $<10^4$~cm$^{-3}$ and shock velocities of $<20$~km s$^{-1}$, where the dependence on the shock velocity is much larger than that on the density \citep{cha91}, this result suggests that the velocities of the molecular shocks are likely to be almost the same along the southern shell of the SNR. At the southern edge around $({\alpha}_{\rm J2000}$, ${\delta}_{\rm J2000})$=$(06^{\rm h}17^{\rm m}25^{\rm s}, +22{\degr}25{\arcmin}00{\arcsec})$, however, the corresponding excitation temperature is ${\sim}4000$~K. Previous studies find that the pulsar wind nebula is present around this region \citep[e.g.][]{olb01}. Hence the molecular gas therein may be heated dominantly by hard radiation fields from the pulsar wind nebula, rather than by shocks, resulting in the excitation temperature higher than that of the shocked molecular gas. Here we assume the gas density in the southern edge region to be significantly lower than $10^4$~cm$^{-3}$, above which the H$_2$ 1--0 S(1) and 2--1 S(1) lines should be thermalized \citep[e.g.][]{ste89}.

\section{Discussion} \label{sec:dis}

\subsection{Properties of the molecular cloud interacting with cosmic rays}

\begin{figure}
%\centering
%\includegraphics{fig2.eps}
\plotone{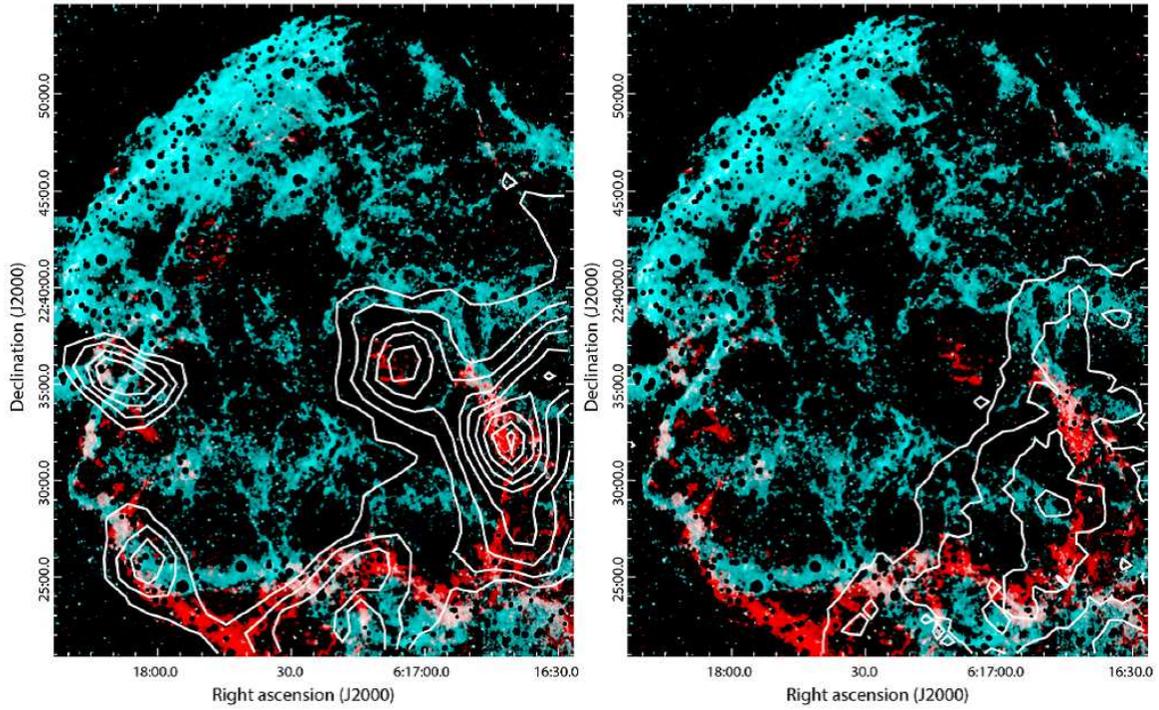}
\caption{Pseudo-color images of IC~443 (blue: [\ion{Fe}{2}] $1.644~{\micron}$, red: H$_2$ 1--0 S(1)), where point sources are removed and the contours of $^{12}$CO ($J=1$--$0$) in a velocity range from $-18$ to $+10$~km~s$^{-1}$ (left; Yoshiike et al. 2020, in prep.) and the TeV $\gamma$ rays \citep[right;][]{acc09} are superimposed. The spatial resolution of the [\ion{Fe}{2}] and H$_2$ maps is as in Figure~\ref{fig:h2}.
\label{fig:h2-fe}}
\end{figure}

From Figure~\ref{fig:h2-fe}, we can see that the $\gamma$-ray emission is significantly detected in the southwestern region of IC~443. This suggests that many cosmic-ray protons interact with the ISM protons not only in post-shock regions but also in pre-shock regions, producing strong $\gamma$-ray emission through the decay of neutral pions. Indeed such a picture is confirmed by detailed analysis of the $\gamma$-ray spectrum of IC~443 \citep{ack13}. As the shock velocities in the southern shell are likely to be almost the same (see Figure~\ref{fig:h2temp}), cosmic-ray protons may not be preferentially accelerated in the southwestern region. Hence the strong $\gamma$-ray emission is likely due to the presence of large amounts of the ISM protons. Indeed Figure~\ref{fig:h2-fe} shows the dominance of the H$_2$ line emission in the southwestern region, indicating that shocks are heavily decelerated by the dense ISM, which can be targets of cosmic-ray protons.

Assuming a molecular cloud size of $2$~pc in the southwestern region and a homogeneous distribution of molecular gas, Yoshiike et al. (2020, in prep.) derive the ISM proton density of $680$~cm$^{-3}$. Here we calculate the penetration depth of cosmic-ray protons in the molecular cloud using the following equation \citep{ino12}:
\begin{equation}
l_{\rm pd}=0.1\ {\eta}^{1/2}\ \left(\frac{E}{10\ {\rm TeV}}\right)^{1/2}\ \left(\frac{B}{100\ {\mu}{\rm G}}\right)^{-1/2}\ \left(\frac{t_{\rm age}}{10^3\ {\rm yr}}\right)^{1/2}\ {\rm pc},
\label{eq1}
\end{equation}
where $\eta$, $E$, $B$, and $t_{\rm age}$ are the degree of magnetic field fluctuations, the energy of cosmic-ray protons, the magnetic field strength, and the age of an SNR, respectively. $\eta$ is defined as $B^2/{\delta}B^2$, which is a measure of the efficiency of the cosmic-ray acceleration, and we adopt $\eta=1$ typical for the shock-cloud interaction \citep[e.g.][]{uch07}. We assume $E=10$~GeV as the hadronic $\gamma$-ray spectrum of IC~443 has its peak around $1$~GeV \citep{ack13} and typically ${\sim}10{\%}$ of the cosmic-ray proton energy is consumed in the $\gamma$-ray emission. The magnetic field strength in the Galactic molecular gas is calculated by
\begin{equation}
B=10\left(\frac{n}{300\ {\rm cm}^{-3}}\right)^{0.65}\ {\mu}{\rm G},
\label{eq2}
\end{equation}
where $n$ is the proton density \citep{cru10}. Using $n=680\ {\rm cm}^{-3}$ estimated from the CO observation of IC~443 (Yoshiike et al. 2020, in prep.), we derive $B=17~{\mu}$G. By combining those parameters and $t_{\rm age}=10^4$~yr obtained for IC~443 \citep{olb01}, $l_{\rm pd}$ is estimated to be ${\sim}0.02$~pc, which is notably smaller than the molecular cloud size of $2$~pc. This result suggests that cosmic-ray protons cannot interact with large amounts of the ISM protons in the southwestern region. However the above cloud size may not be accurate due to the limited spatial resolution of the CO observation by Yoshiike et al. (2020, in prep.). One possibility to explain the strong $\gamma$ rays is the presence of clumpy molecular clouds, which can efficiently interact with many cosmic-ray protons. For instance, we can see that the number of the ISM protons interacting with cosmic rays for clumpy clouds with $n=6800$~cm$^{-3}$ is five times larger than that for uniform clouds with $n=680$~cm$^{-3}$ using Equations~(\ref{eq1}) and (\ref{eq2}), where we assume that the clumpy and uniform clouds have the same cross section. Indeed recent observations with the Atacama Large Millimeter/submillimeter Array (ALMA) reveal that clumpy molecular clouds with its size less than $1$~pc are present in other SNRs \citep[e.g.][]{yam18,san19}.

\subsection{Properties of the ISM interacting with cosmic rays as inferred from the [\ion{Fe}{2}] and H$_2$ maps}

\begin{figure}
\centering
\includegraphics[scale=0.7]{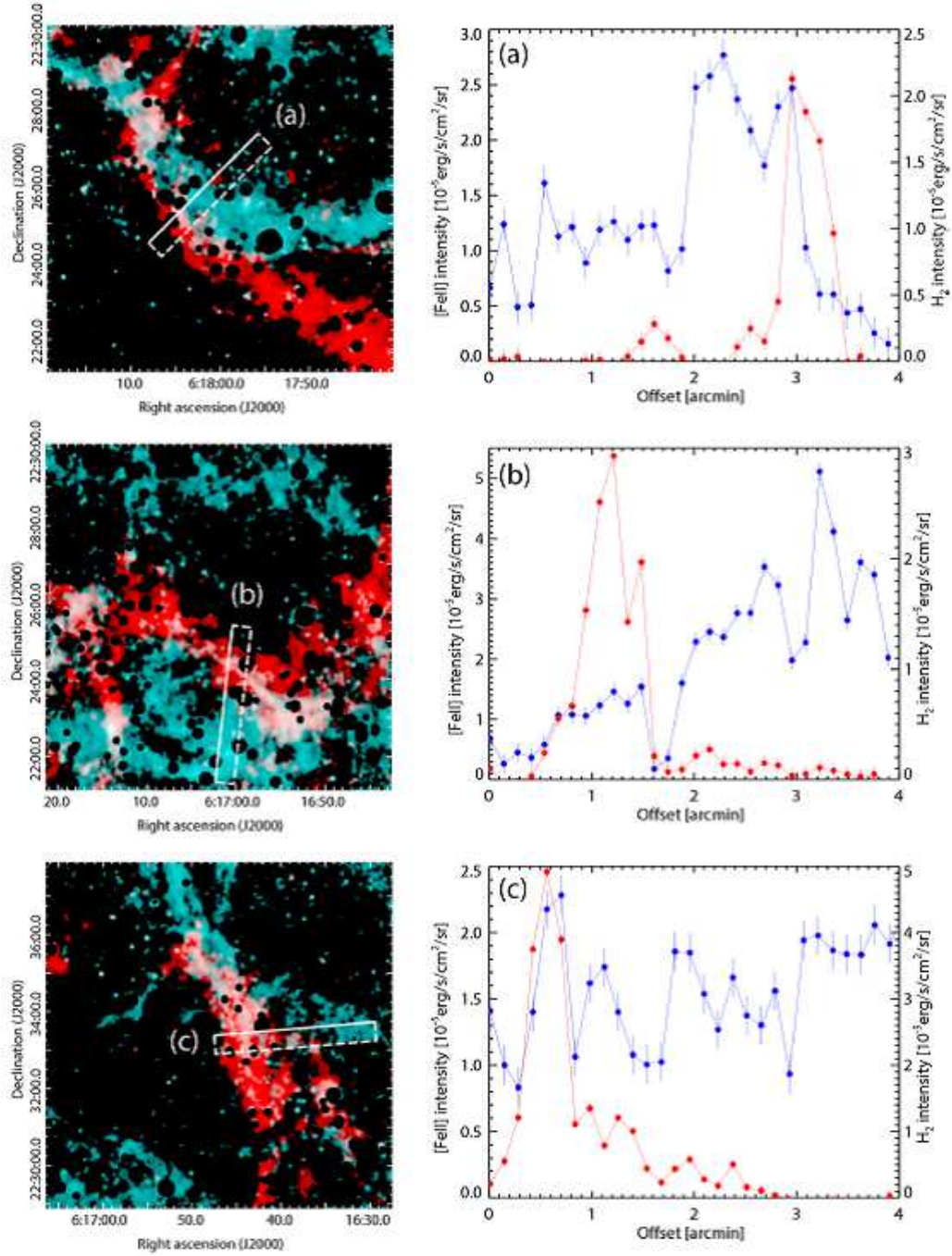}
\caption{Left panels: enlarged pseudo-color images of the southeastern (top) and southwestern (middle and bottom) regions of IC~443 (blue: [\ion{Fe}{2}] $1.644~{\micron}$, red: H$_2$ 1--0 S(1)). The spatial resolution of the [\ion{Fe}{2}] and H$_2$ maps is as in Figure~\ref{fig:h2}. Right panels: cutting profiles of white rectangular regions denoted as (a), (b), and (c) in the left panels. The image pixels where point sources are removed are not used in creating the plots. [\ion{Fe}{2}] and H$_2$ line intensities are plotted in blue and red, respectively, from the center to the outside of the shell.
\label{fig:profile}}
\end{figure}

The right panels of Figure~\ref{fig:profile} show the cutting profiles of the [\ion{Fe}{2}] and H$_2$ line emissions in the southeastern and southwestern regions of IC~443 along the directions denoted in the left panels. The figure demonstrates that the [\ion{Fe}{2}] shell is located inside the H$_2$ shell in the southeastern region, implying that the shock is decelerated by a large dense cloud, and thus fast shocks cannot propagate beyond the cloud, although a projection effect is potentially present in the scenarios suggested here and hereafter. On the other hand, the [\ion{Fe}{2}] shell is distributed well beyond the H$_2$ shell in the southwestern region, indicating that parts of the cloud may be clumpy, and thus fast shocks can propagate through the cloud, while shocks are decelerated in a dense part of the cloud. Indeed \citet{shi11} suggest that shocks may interact with both rarefied clouds and dense clouds around the shell shown in Figures~\ref{fig:profile}(c)  by analyzing the H$_2$ population diagram obtained from the spectral observations in the near- and mid-IR wavelengths.

The above result also suggests that the ISM is likely to be clumpy if the [\ion{Fe}{2}] and H$_2$ shells are spatially anti-correlated to each other as inferred from Figure~\ref{fig:profile}(b). In order to characterize such a case quantitatively, we calculate Peason's linear correlation coefficient $r$ between the [\ion{Fe}{2}] and H$_2$ line intensities over the entire remnant. The presence of dense clumpy clouds is expected to show a large negative $r$ as dense clumpy clouds are bright in H$_2$ but faint in [\ion{Fe}{2}] line emission, while rarefied clouds surrounding clumps are bright in [\ion{Fe}{2}] but faint in H$_2$ line emission. We estimate $r$ based on the method used in \citet{koo16} with our [\ion{Fe}{2}] $1.644~{\mu}$m and H$_2$ 1--0 S(1) line intensity maps as follows: we first define a $1\farcm5\ {\times}\ 1\farcm5$ region centered at each pixel. Then we calculate $r$ between the [\ion{Fe}{2}] and H$_2$ line intensities for each region using the pixels where either [\ion{Fe}{2}] or H$_2$ line emission is detected with $S/N>5$. In this procedure, we adopt the size of the region to calculate $r$ of $1\farcm5\ {\times}\ 1\farcm5$ because the width of the [\ion{Fe}{2}] and H$_2$ shells are less than that size and it reasonably characterizes the spatial difference between the distributions of the [\ion{Fe}{2}] and H$_2$ line emissions.

\begin{figure}
\centering
\includegraphics[scale=0.7]{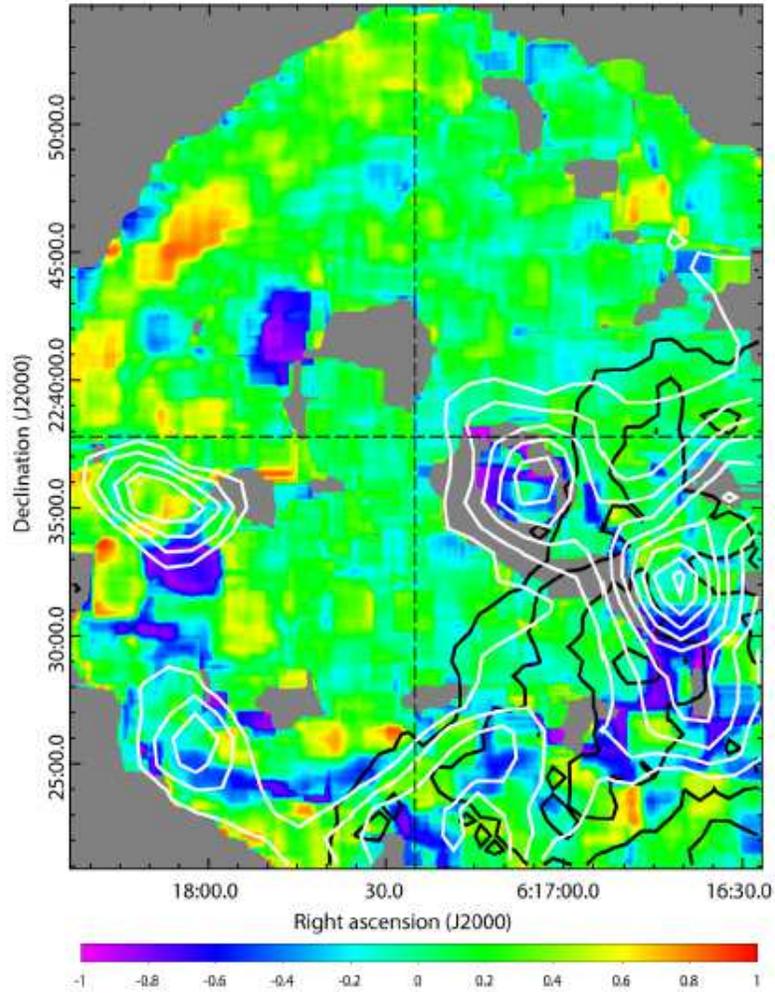}
\caption{Map of the linear correlation coefficients between the [\ion{Fe}{2}] $1.644~\mu$m and H$_2$ 1--0 S(1) line intensities, with the contours of $^{12}$CO ($J=1$--$0$) in a velocity range from $-18$ to $+10$~km~s$^{-1}$ (white; Yoshiike et al. 2020, in prep.) and the TeV $\gamma$ rays \citep[black;][]{acc09}. Black dotted lines show the boundaries of the four quadrants of the map (see Figure~\ref{fig:rhist}).
\label{fig:rmap}}
\end{figure}

\begin{figure}
\centering
\includegraphics[scale=0.5]{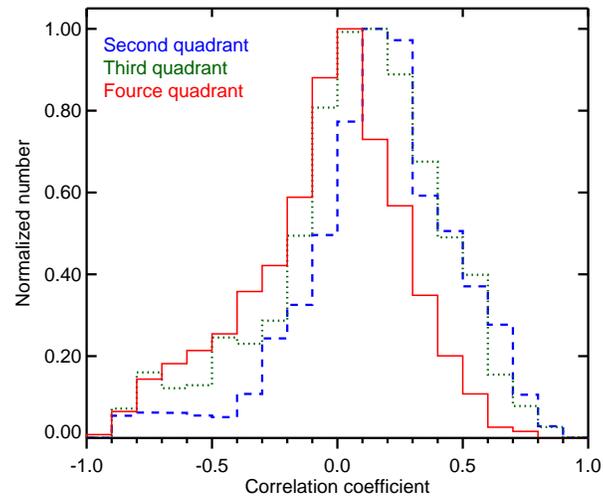}
\caption{Histogram of the linear correlation coefficients between the [\ion{Fe}{2}] $1.644~\mu$m and H$_2$ 1--0 S(1) line intensities. Blue dashed, green doted, and red solid lines show the linear correlation coefficients in the second, the third, and the fourth quadrants, respectively, of the map in Figure~\ref{fig:rmap}.
\label{fig:rhist}}
\end{figure}

Figure~\ref{fig:rmap} shows the resultant map of $r$ between the [\ion{Fe}{2}] and H$_2$ line intensities. The figure shows that the southwestern region has relatively large negative $r$. More quantitatively, Figure~\ref{fig:rhist} shows the histograms of the second, third, and fourth quadrants of the $r$ map, demonstrating that the fourth quadrant covering the southwestern region, where the strong $\gamma$ rays are detected, tends to have a large negative $r$, again suggesting that the ISM clouds in the southwestern region may be clumpy. There are some regions where the values of $r$ become negative in the second and third quadrants, but no $\gamma$-ray emission is detected there. These regions contain molecular clouds considerably less than the southwestern region (see Figure~\ref{fig:rmap}), which may explain that the $\gamma$-ray emission is strong only in the southwestern region. Overall, our near-IR observation suggests the presence of the clumpy ISM in the southwestern region, where cosmic-ray protons may efficiently interact with the dense ISM protons to produce the strong $\gamma$-ray emission. In addition, cosmic rays may be efficiently accelerated due to the effect of the clumpy ISM in the same region \citep[e.g.][]{ino12,san15,mar18}.

From Figures~\ref{fig:rmap} and \ref{fig:rhist}, it is also seen that the northeastern region generally has a large positive $r$, which means that the [\ion{Fe}{2}] and H$_2$ line intensities are correlated with each other. This suggests that H$_2$ may have been reformed on grain surfaces and released from the grains in a high vibrational state after the passage of dissociative fast shocks \citep{hol89}. Indeed \citet{kok13} detect a large amount of dust associated with the SNR along the northeastern shell, where the dust may assist the H$_2$ reformation. Such a formation-pumped H$_2$ is expected to show lower H$_2$ 1--0 S(1)/2--1 S(1) line ratios than collisionally-excited H$_2$, although the H$_2$ 2--1 S(1) emission is not significantly detected in the northeastern region in the present study. More sensitive near-IR H$_2$ observation is needed to verify this scenario.

A projection effect should be present to some degree, and accordingly further supporting evidence is needed to confirm the scenario. For instance, the observed spatial (anti-)correlation between the [\ion{Fe}{2}] and H$_2$ line emissions may be due to an accidental overlap of the line emissions from different clouds present in a line of sight. In order to reveal whether the ISM interacting with cosmic-ray protons in IC~443 is truly clumpy or not, we plan to conduct CO observations of the SNR with ALMA, the result of which will be reported in a separate paper.

\section{Conclusions} \label{sec:con}
We have investigated the properties of the ISM interacting with shocks around the Galactic supernova remnant IC~443, using the results of near-IR [\ion{Fe}{2}] and H$_2$ line mapping with the IRSF/SIRIUS. In the present study, we newly performed H$_2$ 1--0 S(1) and 2--1 S(1) line mapping with the narrow-band filters tuned for these lines. Covering the entire remnant ($30{\arcmin}\ {\times}\ 35{\arcmin}$), our observation presents the large-area H$_2$ line intensity maps of IC~443, which reveal that the H$_2$ emission is mainly distributed in the southern shell. On the other hand, the [\ion{Fe}{2}] line emission is detected all over the remnant, suggesting that slow and fast shocks propagate through the southern region and the entire remnant, respectively. With the H$_2$ 1--0 S(1)/2--1 S(1) line ratio, we estimated the H$_2$ excitation temperature in the southern shell, finding that the temperature is almost constant along the shell. As the temperature of shocked gas is a measure of shock velocities, this result implies that the velocities of the molecular shocks are likely to be almost the same along the southern shell.

In the southwestern region of IC~443, the strong TeV $\gamma$ rays of cosmic-ray proton origins are detected. As the shock velocities in the southern shell are likely to be almost the same, cosmic-ray protons may not be preferentially accelerated in the southwestern region, and the strong $\gamma$-ray emission is likely due to the presence of large amounts of the ISM protons. Indeed the H$_2$ line emission is relatively strong compared to the [\ion{Fe}{2}] line emission in the southwestern region,   supporting the scenario that the $\gamma$-ray emission is likely to be produced through a heavy interaction between cosmic-ray protons and the dense ISM.

Our near-IR observation reveals that the [\ion{Fe}{2}] shell is distributed well beyond the H$_2$ shell in the southwestern region, indicating that parts of the cloud may be clumpy, and thus fast shocks can propagate through the cloud, while shocks are decelerated in a dense part of the cloud. In addition, the [\ion{Fe}{2}] and H$_2$ line intensities tend to be anti-correlated in the same region. As dense clumpy clouds and rarefied clouds surrounding clumps are expected to be bright only in H$_2$ and [\ion{Fe}{2}] emissions, respectively, the above result further supports the presence of clumpy molecular clouds. Such a clumpy molecular clouds, which may not be spatially resolved in the radio observation of IC~443 conducted so far, may assist cosmic-ray protons to efficiently interact with large amounts of the ISM protons. 

\vspace{\baselineskip}

The IRSF project was financially supported by the Sumitomo foundation and Grants-in-Aid for Scientific Research on Priority Areas (A) (Nos. 10147207 and 10147214) from the Ministry of Education, Culture, Sports, Science and Technology (MEXT). The operation of IRSF is supported by Joint Development Research of National Astronomical Observatory of Japan, and Optical Near-Infrared Astronomy Inter-University Cooperation Program, funded by the MEXT of Japan.

\clearpage

\begin{deluxetable}{cccccc}
\tablecaption{Summary of the present IRSF observations \label{table:obs}}
\tablewidth{0pt}
\tablehead{
\colhead{${\lambda}_{\rm center}$ ($\micron$)} & \colhead{Transition} & \colhead{${\Delta}{\lambda}_{\rm eff}$\tablenotemark{a} ($\micron$)} &
\colhead{Exposure (s)} & \colhead{NDI\tablenotemark{b}} & \colhead{Date}
}
\startdata
2.121 & 1--0 S(1)  & 0.033 & 60 & 10 & Nov 2012, Feb 2013\\
2.247 & 2--1 S(1)  & 0.032 & 30 & 15 & Mar 2014, Mar 2019\\ 
\enddata
\tablenotetext{a}{Effective band width calculated from ${\int}S({\lambda})d{\lambda}=T_{\lambda}{\Delta}{\lambda}_{\rm eff}$, where $S({\lambda})$ is a filter response curve, and $T_{\lambda}$ is a throughput at ${\lambda}_{\rm center}$.}
\tablenotetext{b}{Number of dithered images. The total exposure times for each field are $600$ and $450$ seconds for the 1--0 S(1) and 2--1 S(1) lines, respectively.}
\end{deluxetable}

\clearpage




%\begin{thebibliography}{}

%\end{thebibliography}

\bibliography{ref}



\end{document}